\documentclass[shortnote,twocolumn]{jpsj3}

\title{
  Weak Value in Wave Function of Detector
}

\author{
  Kouji Nakamura\thanks{E-mail address: kouji.nakamura@nao.ac.jp
  } 
}
\inst{
  Optical and Infrared Astronomy Division,
  National Astronomical Observatory of Japan,\\
  Osawa 2-21-1, Mitaka, Tokyo, 181-8588, Japan. 
}
\abst{
  A simple theoretical formula to derive the weak value from
  the wave function of the detector after the postselection with
  the initial Gaussian profile is proposed.
  We apply this formula for the weak value to an experiment of
  the realization of the weak measurement by the optical
  polarization and obtain the weak value for any pre- and
  post-selections.
  This formula automatically includes the interference effect
  which is necessary to yields the weak value as an outcome of
  the weak measurement and it will be useful to theoretical
  discussion on weak measurements.
}

\kword{
  weak value, weak measurement, formula for weak value, wave
  function, detector
}

\begin{document}
\maketitle

``Weak measurement'' and its outcome, ``weak value'', are
topical subjects in the recent development of quantum
mechanics\cite{Y.Aharonov-D.Z.Albert-L.Vaidman-1988,I.M.Duck-P.M.Stevenson-E.C.G.Sudarshan-1989,Y.Aharonov-L.Vaidman-1990,Y.Aharonov-L.Vaidman-2007}.
This weak measurement is not only a theoretical concept but also
applied to many
experiments\cite{N.W.M.Ritchie-J.G.Story-R.G.Hulet-1991,P.B.Dixon-D.J.Starling-A.N.Jordan-J.C.Howell-2009}. 
In particular, through the weak measurement, we have an
amplification of the outcome which is much larger than the
eigenvalue spectrum of the observable.
Therefore, the properties of weak measurements and their
applications are interesting issues in recent physics.


In the weak measurement, the total system consists of the
``system'' and the ``detector''.
The system is a quantum system and we want to measure an
observable of this system.
The detector is also a quantum system that interacts
with the system for measuring an observable of the system.
In some cases, the wave function of the detector after the weak
measurement has a complicated form, even when we know the
explicit analytical expression of the wave
function\cite{Y.Aharonov-L.Vaidman-2007}. 
In such cases, it is tedius to derive the outcome, i.e., ``weak
value'' from complicated wave functions.
Therefore, a simple formula to derive the weak value in the
explicit wave function is useful.
In this short note, we propose a simple theoretical formula for
deriving the weak value from the explicit wave function of the
detector after the weak measurement.


The process of the weak measurement consists of a sequence
of four measurements\cite{Y.Aharonov-D.Z.Albert-L.Vaidman-1988}. 
The first three processes of these four measurements are called
``preselection'', ``weak measurement'', and ``postselection''.
The final one is the measurement of the pointer variable of the
detector by any type of applicable measurement in quantum
mechanics.


First, we prepare the initial state $|\Psi\rangle$ of the system
through the projection measurement at $t=t_{1}$, which is called 
``preselection''.
We also prepare the initial state of the detector
$|\Psi^{D}_{in}\rangle$, and the total system is described by 
the state $|\Psi\rangle\otimes|\Psi^{D}_{in}\rangle$.
In this note, we assume that the initial wave function
$\Psi_{in}^{D}:=\langle Q|\Psi_{in}^{D}\rangle$ of the
detector is a Gaussian distribution 
\begin{eqnarray}
  \label{eq:K.Nakamura-initial-state-MD-wave}
  \Psi_{in}^{D}
  =
  \left(\frac{\beta}{\pi}\right)^{1/4}
  \exp\left(-\frac{\beta Q^{2}}{2}\right), 
\end{eqnarray}
where $Q$ is the pointer variable of the detector and
$\beta^{-1/2}$ is the width of this Gaussian distribution.


Second, after the preselection at $t=t_{1}$, the switch of the
interaction between the system and the detector is turned on and
the detector measures the observable $C$ of the system through
this interaction.
Typically, this interaction is described by the von Neumann
measurement\cite{J.von-Neumann-1932} with the interaction
Hamiltonian
\begin{eqnarray}
  \label{eq:K.Nakamura-vonNeumann-interaction}
  H = g(t)PC,
\end{eqnarray}
where $P$ is the conjugate momentum to the pointer
variable $Q$ and the function $g(t)$ is the switch of the
interaction with the support $t\in[t_{1},t_{2}]$, $t_{1}<t_{2}$; 
$g(t)=0$ outside of this support.
In some papers, the ``weak'' of the weak measurement means
$g(t)$ is sufficiently small and then the disturbance to the
system from the measurement by the detector is negligible.
However, this coupling constant is completely determined when we
specify the concrete system and the detector, and we cannot
control it after this specification.
Therefore, in this note, we choose $\int_{t_{1}}^{t_{2}}g(t)=1$.
Instead, we choose $\beta$ so that $\beta^{-1/2}$ is
sufficiently large, as the original proposal of the weak
measurement\cite{Y.Aharonov-D.Z.Albert-L.Vaidman-1988}.
$\beta$ is controllable through the preparation of the initial
state of the detector, and this choice yields negligible
disturbance to the system through the measurement, and then, so
called ``weak measurement'' is accomplished.


Next, we perform ``postselection'', which is the projection
measurement of the system to the state $|\Phi\rangle$ at
$t=t_{2}$.
It is important to note that we choose the postselected state
$|\Phi\rangle$ such that $|\Phi\rangle$ is not orthogonal to the 
preselected state $|\Psi\rangle$ of the system
($\langle\Phi|\Psi\rangle\neq 0$) but nearly orthogonal.
Then, the state $|\Psi^{D}_{out}\rangle$ of the detector after
the postselection is given by
\begin{eqnarray}
  \label{eq:K.Nakamura-wave-function-of-MD-after-postselection}
  |\Psi^{D}_{out}\rangle
  =
  \langle\Phi|\exp\left(-i\int_{t_{1}}^{t_{2}}dt'H(t')\right)|\Psi\rangle\otimes|\Psi^{D}_{in}\rangle.
\end{eqnarray}
As shown by Aharonov and
Vaidmann\cite{Y.Aharonov-L.Vaidman-1990}, the wave function 
$\Psi^{D}_{out}:=\langle Q|\Psi^{D}_{out}\rangle$ of the
detector after postselection is given by
\begin{eqnarray}
  \Psi^{D}_{out}
  &\propto&
  \exp\left\{
    - \frac{\beta}{2}(Q-C_{w})^{2}
  \right\}
  \nonumber\\
  &&
  +
  e^{-\frac{\beta Q^{2}}{2}}
  \sum_{n=2}^{\infty} \frac{\left(C^{n}\right)_{w} - (C_{w})^{n}}{n!\sqrt{\pi}} 
  \left(-\sqrt{2\beta}i\right)^{n}
  \nonumber\\
  &&
  \quad\quad\quad
  \times
  \int_{-\infty}^{\infty}dx
  \left(x + \frac{i\beta^{1/2}Q}{\sqrt{2}}\right)^{n}
  e^{-x^{2}}
  ,
  \label{eq:kouchan-note-8-10}
\end{eqnarray}
where $C_{w}$ is called a ``weak value'' and is defined by  
\begin{eqnarray}
  \label{eq:Y.Aharonov-L.Vaidman-2001-1.1-21}
  C_{w} := \frac{
    <\Phi|C|\Psi>
  }{
    <\Phi|\Psi>
  }
\end{eqnarray}
and $(C^{n})_{w}:=\langle\Phi|C^{n}|\Psi\rangle/\langle\Phi|\Psi\rangle$.


If the interaction between the system and the detector is really
``weak'', this sequence of the measurements yields the 
shift of the pointer variable $Q$ of the detector by the weak
value $C_{w}$.
This can be easily seen from eq.~(\ref{eq:kouchan-note-8-10}).
In our ``weak'' measurement, $\beta^{-1/2}$ is sufficiently
large, i.e., the initial wave function of the detector has a
broad profile.
Then the second term in the right-hand side of
eq.~(\ref{eq:kouchan-note-8-10}) is negligible.
Because of the weakness of the measurement, a single weak
measurement of the detector may not yield this weak value, i.e.,
the measurement is imprecise.
However, by performing the weak measurement on an ensemble of
$N$ identical systems, this imprecision is improved by the
factor $\sqrt{N}$.
This is the success of the weak measurement.
Here, we must note that, according to the value of
$\beta^{-1/2}$ and the pre- and postselections, the weak
measurement may fail to succeed and we may not obtain the weak
value as the shift in $\Psi^{D}_{out}$.
In this case, the second term on the right-hand side of
eq.~(\ref{eq:kouchan-note-8-10}) is not negligible.




Now, we propose a simple formula for calculating the weak value 
from the wave function of the detector after the postselection
in weak measurement.
To extract the weak value from the wave function
(\ref{eq:kouchan-note-8-10}), the overall normalization of the 
wave function is irrelevant.
To extract the information that is independent of this overall
normalization, we consider the partial derivative of the
logarithm of eq.~(\ref{eq:kouchan-note-8-10}) with respect to
$Q$.
Furthermore, from the order counting of $\beta$, we can expect
that the derivative of the logarithm of
eq.~(\ref{eq:kouchan-note-8-10}) with respect to $\beta$ is
related to the weak value $C_{w}$.
Actually, we can easily show the formula
\begin{eqnarray}
  \label{eq:kouchan-formula-main-result}
  \left.\frac{\partial}{\partial\beta}\left[
      \left.
        \frac{\partial}{\partial
          Q}\ln\left(\Psi^{D}_{out}(Q)\right)
      \right|_{Q=0}\right]
  \right|_{\beta=0}
  =
  C_{w}
  .
\end{eqnarray}
Through this formula, if the initial wave function of the
detector is Gaussian, we can extract the weak value from the
wave function of the detector after the postselection, 
even if the wave function has a complicated form.
In eq.~(\ref{eq:kouchan-formula-main-result}), $\beta=0$
corresponds to the weak limit of the measurement, i.e., the
second term in eq.~(\ref{eq:kouchan-note-8-10}) does not affect
to the calculation using
eq.~(\ref{eq:kouchan-formula-main-result}).




Next, as an example, we explicitly derive the weak value in a
concrete experiment using the formula
(\ref{eq:kouchan-formula-main-result}).
The experiment that we consider here is an optical analog of 
the Stern-Gerlach spin-$\frac{1}{2}$ experiment proposed by Duck
et al.\cite{I.M.Duck-P.M.Stevenson-E.C.G.Sudarshan-1989}, and 
carried out by Ritchie et
al.\cite{N.W.M.Ritchie-J.G.Story-R.G.Hulet-1991}.
In this analog, the beam of spin-$\frac{1}{2}$ particles is
replaced by a Gaussian-mode laser beam and the preselection and 
postselection Stern-Gerlach magnets are replaced by optical
polarizers.
The weak measurement is performed using a
birefringent-crystalline quartz plate that spatially separates
the two orthogonal polarizations of the laser radiation by a
distance much less than the Gaussian beam waist of the laser
beam.


Now, we consider the laser to be propagating along the $z$
direction and to be linearly polarized at an angle $\alpha$ with
respect to the $x$ axis as the preselection.
The electric-field vector of this radiation is described by 
\begin{eqnarray}
  \label{eq:N.W.M.Ritchie-J.G.Story-R.G.Hulet-1991-3}
  {\bf E}_{i} = E_{0}\exp\left(-\frac{\beta(x^{2}+y^{2})}{2}\right)
  \left(
    \cos\alpha {\bf \hat{x}} + \sin\alpha {\bf \hat{y}}
  \right),
\end{eqnarray}
where $\beta^{-1/2}$ is the beam waist.
In this experiment, the system in the weak measurement is the
polarization of the laser beam and the detector in the
weak measurement is the transverse distribution of the laser
beam.
Therefore,
eq.~(\ref{eq:N.W.M.Ritchie-J.G.Story-R.G.Hulet-1991-3}) shows
that the initial state of the detector has the Gaussian
distribution.
The laser beam is incident on a plane-parallel uniaxial
birefringent plate whose optic axis is aligned with the $x$
axis.
The plane of the plate includes the $x$ axis and is rotated from
the $y$ axis by angle $\theta$.
The birefringent plate performs a weak measurement by spatially
separating the two orthogonal linear-polarization components of
the field, corresponding to the ordinary and extraordinary
rays\cite{E.Hecht-1987}, by a distance $a$ which is determined
by $\theta$ and is small compared with $\beta^{-1/2}$.
After this weak measurement, the postselection is performed by a
polarizer aligned at angle $\alpha'$ with respect to the
$x$-axis.
Then, the wave function of the detector after this postselection
is given by
\begin{eqnarray}
  \Psi^{D}(y)
  &\propto& 
  \cos\alpha\cos\alpha' e^{-\beta(y+a)^{2}/2}
  \nonumber\\
  && \quad
  +
  \sin\alpha\sin\alpha' e^{-\beta y^{2}/2}
  \label{eq:N.W.M.Ritchie-J.G.Story-R.G.Hulet-1991-5}
  .
\end{eqnarray}


From this wave function
(\ref{eq:N.W.M.Ritchie-J.G.Story-R.G.Hulet-1991-5}) of the
detector, we can derive the weak value through formula
(\ref{eq:kouchan-formula-main-result}):
\begin{eqnarray}
  \label{eq:N.W.M.Ritchie-weak-value-exact}
  C_{w} = - \frac{a}{1 + \tan\alpha\tan\alpha'}.
\end{eqnarray}
If we choose $\alpha=\pi/4$ and $\alpha'=\alpha+\pi/2+\epsilon$
with $\epsilon\ll 1$, we have
$C_{w}=-\frac{1}{2}a(1+\tan\epsilon)\cot\epsilon$, which
coincides with the estimated weak value reported by Ritchie et 
al.\cite{N.W.M.Ritchie-J.G.Story-R.G.Hulet-1991}.


In calculation (\ref{eq:N.W.M.Ritchie-weak-value-exact}),
the interference effect between two Gaussian profiles in
eq.~(\ref{eq:N.W.M.Ritchie-J.G.Story-R.G.Hulet-1991-5}) is
automatically taken into account through the derivative of the 
logarithm of the wave function.
The coexistence of the two terms on the right-hand side of
eq.~(\ref{eq:N.W.M.Ritchie-J.G.Story-R.G.Hulet-1991-5}) is
essential to the nontrivial weak value
(\ref{eq:N.W.M.Ritchie-weak-value-exact}).
Thus, through the calculation of the weak value by formula
(\ref{eq:kouchan-formula-main-result}), we can easily see that
the weak value is realized through the interference of these two
Gaussian profles in this experiment.




In summary, we proposed a theoretical formula for deriving the
weak value from the detector wave function after the
postselection.
We must emphasize that our formula
(\ref{eq:kouchan-formula-main-result}) automatically includes
the interference effect, which is necessary for the success of
the weak measurement, originating from the derivative of the
logarithm of the wave function in
eq.~(\ref{eq:kouchan-formula-main-result}). 
Therefore, we may say that our formula
(\ref{eq:kouchan-formula-main-result}) is a natural form of the 
weak value and this formula will be useful when we derive the
weak value from complicated analytic wave functions of the
detector.


The author deeply thanks Professor Masa-katsu Fujimoto for
giving the author the opportunity to research this topic.
This work is carried out as part of the research assistant work
for Profssor Masa-katsu Fujimoto.
Furthermore, the author also acknowledges valuable conversations
with Dr. Atsushi Nishizawa.


\end{document}